\documentstyle[12pt]{article}
\begin{document}
\def\thefootnote{\fnsymbol{footnote}}
\begin{flushright}
KANAZAWA-02-11  \\ 
July, 2002
\end{flushright}
\vspace*{2cm}
\begin{center}
{\LARGE\bf  D-term inflation and neutrino mass}\\
\vspace{1 cm}
{\Large Daijiro Suematsu}
\footnote[1]{e-mail: suematsu@hep.s.kanazawa-u.ac.jp}
\vspace*{1cm}\\
{\it Institute for Theoretical Physics, Kanazawa University,\\
        Kanazawa 920-1192, Japan}\\    
\end{center}
\vspace{1cm}
{\Large\bf Abstract}\\  
We study a $D$-term inflation scenario in a model extended from the 
minimal supersymmetric standard model (MSSM) by two additional 
abelian factor groups focussing on its particle physics aspects. 
Condensates of the fields related to the inflation can naturally 
give a possible solution to both the $\mu$-problem in the MSSM and 
the neutrino mass through their nonrenormalizable couplings to the 
MSSM fields. Mixings between neutrinos and neutralinos are also 
induced by some of these condensates. Small neutrino masses are 
generated by a weak scale seesaw mechanism as a result of these 
mixings. Moreover, the decay of the condensates may be able to cause 
the leptogenesis. Usually known discrepancy between both values of 
a Fayet-Iliopoulos $D$-term which are predicted by the COBE 
normalization and also by an anomalous U(1) in the weakly-coupled 
superstring might be reconciled.
\newpage
\setcounter{footnote}{0}
\def\thefootnote{\arabic{footnote}}
\section{Introduction}
Recent observations in both astrophysics and 
particle physics provide crucial informations to astro-particle physics. 
Observation of the cosmic microwave background radiation (CMB) gives 
a strong support for the inflation \cite{cobe}. 
Moreover, precise estimations of the spectral index of the density
perturbations based on the recent
observations strictly restrict the form of the inflaton potential.
Particle physics models seem to be 
required to have a structure to induce the inflation consistent with
these observations in it.
On the other hand, observations of the atmospheric and solar neutrinos at 
Super-Kamiokande suggest the existence of nonzero neutrino masses 
and this fact requires some
modification of the standard model (SM) \cite{sk}. 
From the theoretical point of view of particle physics, moreover,
supersymmetrization of the SM is considered to be the most 
promising candidate to solve the gauge hierarchy problem, 
although we have no direct evidence for it still now.   
Thus at the present stage it seems to be natural to consider the inflation 
based on a suitable supersymmetric model which is motivated from 
a view point of particle physics such as the explanation of, for example, 
the neutrino mass, the $\mu$-problem, the strong $CP$ problem and 
so on \cite{lr}-\cite{dr}.

Various ideas for the inflation \cite{chaotic}-\cite{i-hybrid}
have been extended to the supersymmetric models by now. 
They may be classified into two types of models, that is, 
an $F$-term inflation model and a $D$-term inflation model. 
If we consider the $F$-term inflation, we are always suffered from
the $\eta$-problem which is caused by the supergravity correction 
to an inflaton mass as a result of the supersymmetry breaking during
the inflation \cite{eta}. As far as we do not introduce a special K\"ahler
potential \cite{eta}, there is a large correction to the inflaton mass 
of the order of a Hubble constant and then the inflation cannot occur. 
A $D$-term inflation \cite{bd} has been proposed as a model which 
can escape this $\eta$-problem.  
In the $D$-term inflation model there appear several SM singlet 
fields intimately related to the inflation.
Although they play a crucial role in the inflation phenomena, they
usually have no role in particle physics.
If we can relate their existence to the important problems in particle
physics, for example, the neutrino mass, the $\mu$-problem and so on,
the $D$-term inflation scenario might be much more promising. 
   
In this paper we investigate some particle physics aspects of 
the SM singlet fields involved in the inflation in a certain 
$D$-term inflation model. The model is defined as an extension from the
MSSM by two abelian factor groups 
U(1)$_X\times$U(1)$_A$, which are extended by an additional abelian 
factor group in comparison with the ordinary $D$-term inflation scenario.
We show that the condensates of these singlet fields can be related 
to small neutrino masses and the $\mu$-problem in this model.
Their decay also may induce the leptogenesis.
Moreover, the introduction of this new factor group might be able 
to relax the discrepancy between the scales of the $D$-term predicted 
by the cosmic background explorer (COBE) normalization and by the 
weakly-coupled superstring. 
The paper is organized as follows.
In the next section we discuss various aspects of the inflation 
in this model. In section 3 we study the features as a particle 
physics model, especially, the origin of the neutrino masses 
and the $\mu$-term. The possibility of the leptogenesis is also briefly
discussed. The last section is devoted to the summary.

\section{A model of $D$-term inflation} 
In our model field contents and a gauge structure are motivated 
by the weakly-coupled superstring models based on, for example, $E_6$.
Such kind of the field contents and the gauge structure have been 
suggested to appear often in their effective theories \cite{cceel}.
U(1)$_A$ is assumed to have a Fayet-Iliopoulos (F-I) $D$-term, which might be 
considered to be an anomalous U(1). As fields related to the inflation,
we introduce a gauge singlet chiral superfield $\phi$ and 
two pairs of chiral superfields ($N$, $\bar N$) and ($S$, $\bar S$), 
which are the SM singlet fields but have charges of
U(1)$_X\times$U(1)$_A$. A renormalizable
superpotential constituted by these fields is assumed to be
\begin{equation}
W_1=k\phi N\bar N,
\label{eqa}
\end{equation}
where $k$ is considered to be real and positive by a suitable
redefinition of fields.
A value of $k$ is constrained on the basis of some
observational requirements such as the CMB data and so on.
The potential for the scalar components of these fields can 
be written as\footnote{
In the following discussion we denote a scalar component of the superfield 
by adding a tilde on the same character as the superfield and 
a spinor component by the same notation as them.}
\begin{eqnarray}
V&=&{g_X^2 \over 2}\left[Q_X^{N}\left(\vert\tilde N\vert^2-
\vert\tilde{\bar N}\vert^2\right)+Q_X^{S}\left(\vert\tilde S\vert^2-
\vert\tilde{\bar S}\vert^2\right) \right]^2 \nonumber \\
&+&{g_A^2 \over 2}\left[Q_A^{N}\left(\vert\tilde N\vert^2-
\vert\tilde{\bar N}\vert^2\right)+Q_A^{S}\left(\vert\tilde S\vert^2-
\vert\tilde{\bar S}\vert^2\right)+\xi_A \right]^2 \nonumber \\
&+&k^2\left[\vert\tilde\phi\vert^2\left(\vert\tilde N\vert^2 +
\vert\tilde{\bar N}\vert^2 \right) 
+ \vert\tilde N\tilde{\bar N}\vert^2 \right].
\label{eqb}
\end{eqnarray}
The first two lines are $D$-term contributions and the last line is
an $F$-term contribution coming from eq.~(\ref{eqa}). 
The role of ($N$, $\bar N$) and ($S$, $\bar S$)
in particle physics is determined through the couplings with the MSSM 
matter fields.
These couplings can be neglected in eq.~(\ref{eqb}) since the field
values of the MSSM matter fields are fixed to be zero by the steep potential.
We will come back to the effects of these couplings later and firstly 
we discuss features of the inflation in this model.

As similar to the usual hybrid inflation models, 
several fields concern the inflation in the present model.\footnote{
Although we could consider the model as an chaotic inflation model along the 
$D$-flat direction to which 
several fields are related such as \cite{sy,gk},
it seems to be better to set it up as an hybrid inflation to avoid
extremely small couplings, that is, from the view point of naturalness.}
A potential minimum shifts from the false vacuum to the true one 
around a critical value $\tilde\phi=\tilde\phi_c$ when $\tilde\phi$
varies its value.
If $\tilde\phi$ initially takes a large value such as 
$\vert\tilde\phi\vert\gg \vert\tilde\phi_c\vert$ which is naturally 
expected because of the
flatness of its potential, a minimum is realized at 
\begin{equation}
\tilde N=\tilde{\bar N}=0,  \qquad  
\delta\tilde S^2_i\equiv\vert\tilde S\vert^2-\vert\tilde{\bar S}\vert^2=
-{g_A^2Q_A^S\xi_A\over g_X^2Q_X^{S2}+g_A^2Q_A^{S2}}, 
\label{eqc}
\end{equation}
where we should note that each value of 
$\vert\tilde S\vert$ and $\vert\tilde{\bar S}\vert$ 
cannot be determined here.
At this minimum the potential has a constant value
\begin{equation}
V={1\over 2}g_A^2G\xi_A^2, \qquad G={g_X^2Q_X^{S2}\over g_X^2Q_X^{S2}
+g_A^2Q_A^{S2}}.
\label{eqd}
\end{equation}
Tree level mass of $\tilde\phi$ is zero and the potential is
flat in that direction. However, one-loop effect due to the 
interaction of eq.~(\ref{eqa}) induces an
effective mass in the same way as the models in \cite{bd} since
supersymmetry is broken by the vacuum energy at that time.
In fact, although spinor components of $N$ and $\bar N$ have the same mass
$m_{N,\bar N}=k\vert\tilde\phi\vert$,
the masses of the scalar components $\tilde N$ and $\tilde{\bar N}$ change
differently their values depending on the value of $\tilde\phi$ as  
\begin{eqnarray}
&&m_{\tilde N}^2=k^2\vert\tilde\phi\vert^2+
\left[g_A^2Q_A^N\xi_A+(g_X^2Q_X^SQ_X^N+g_A^2Q_A^SQ_A^N)\delta\tilde S^2 
\right], \nonumber \\
&&m_{\tilde{\bar N}}^2=k^2\vert\tilde\phi\vert^2-
\left[g_A^2Q_A^N\xi_A+(g_X^2Q_X^SQ_X^N+g_A^2Q_A^SQ_A^N)\delta\tilde S^2 
\right]. 
\label{eqf}
\end{eqnarray}
Using these relations, the effective potential \cite{cw} can be written as
\begin{equation}
V(\tilde\phi)={1\over 2}g_A^2G\xi_A^2
+{k^4\vert\tilde\phi_c\vert^4 \over 16\pi^2}
\ln{k^2\vert\tilde\phi\vert^2\over \Lambda^2}.
\label{eqe}
\end{equation}
This potential makes $\tilde\phi$ slowly roll to the critical value 
$\tilde\phi_c$ from its initial one $(\gg\vert\tilde\phi_c\vert)$. 
It results in a $D$-term 
inflation due to the vacuum energy (\ref{eqd}) during that 
slow-roll period.\footnote{If there are 
contributions to the above mentioned
$D$-terms from other fields, a vacuum shift could occur to restore the
supersymmetry and the inflation cannot be induced in that case
\cite{err}.
However, the suitable charge assignment for such fields always make it
possible to avoid such a situation. 
In this paper we consider such a case where their contribution 
to eq.~(\ref{eqb}) can be neglected. }
If charges of the fields are suitably selected, 
either $m_{\tilde{N}}^2$ or
$m_{\tilde{\bar N}}^2$ in eq.~(\ref{eqf}) changes its sign at 
$\tilde\phi=\tilde\phi_c$ and the global minimum
appears in a different point from the one of eq.~(\ref{eqc}).  

When $\vert\tilde\phi\vert <\vert\tilde\phi_c\vert$, we find that  
$m_{\tilde N}^2m_{\tilde{\bar N}}^2<0$
is satisfied so that either $\tilde N$ or $\tilde{\bar N}$
can have a nonzero vacuum expectation value.
If we assume $m_{\tilde{\bar N}}^2<0$, 
the potential minimum shifts from a point described by eq.~(\ref{eqc}) 
to a point with 
$\vert\tilde{\bar N}\vert\not= 0$ and 
$\vert\tilde{N}\vert=0$.    
The inflaton $\tilde\phi$ gets a large mass 
$m_{\tilde\phi}^2=k^2\vert\tilde{\bar N}\vert^2$
so that the slow-roll of $\tilde\phi$ stops and the inflation is
expected to end at 
$\vert\tilde\phi\vert\sim\vert\tilde\phi_c\vert$. 
The critical value $\tilde\phi_c$ is estimated as
\begin{equation}
\vert\tilde\phi_c\vert^2=
{g_A^2\over k^2}\xi_A GF, 
\qquad F={Q_A^NQ_X^S-Q_A^SQ_X^N \over Q_X^S},
\label{eqg}
\end{equation}
where in this derivation we used a value for $\delta\tilde S^2_i$ 
in eq.~(\ref{eqc}).
If we use this formula in eq.~(\ref{eqe}), we can simplify the
expression for $V(\tilde\phi)$ as
\begin{equation}
V(\tilde\phi)={1\over 2}g_A^2G\xi_A^2\left(1+{g_A^2\over 16\pi^2}GF^2
\ln{k^2\vert\phi\vert^2\over \Lambda^2}\right).
\label{eqgg}
\end{equation}
We find that this is a simple modification by factors $G$ and $F$
from the inflaton potential given in \cite{bd}.
The change is caused by the extension of the gauge structure and the
introduction of the new SM singlet fields. 
On the other hand, the vacuum shift also occurs in the sector 
of $(\tilde S, \tilde{\bar S})$
from a state with the value of $\delta\tilde S^2_i$ in eq.~(\ref{eqc}) 
to the global minimum. 
The global minimum is realized at
\begin{equation}
\langle\tilde\phi\rangle=\langle\tilde{N}\rangle=0, \qquad 
\vert\langle\tilde{\bar N}\rangle\vert^2={\xi_A\over F}, 
\qquad
\delta\tilde S^2_f\equiv\vert\langle\tilde{S}\rangle\vert^2
- \vert\langle\tilde{\bar S}\rangle\vert^2
={Q_X^N\over Q_X^S}{\xi_A\over F}.
\label{eqh}
\end{equation}
At this global minimum, $V=0$ and the supersymmetry is restored.
It may be useful to note that the deviations of $\tilde{\bar N}$ 
and $\delta\tilde S^2$ from their true vacuum values could
play a role like an Affleck-Dine (AD) condensates \cite{ad} and be relevant to 
the leptogenesis depending on the coupling of these fields with the
MSSM matter fields. We will come back to this point later.

The superfields $S$, $\bar S$, 
$N$ and $\bar N$ can get masses represented by
\begin{equation}
m_{\tilde N}^2=k^2\vert\langle\tilde{\bar N}\rangle\vert^2, \qquad
{\cal M}^2=2 \left(\begin{array}{ccc}
A\vert\langle\tilde S\rangle\vert^2 &
-A\vert\langle\tilde{\bar S}\rangle\langle\tilde S\rangle\vert&
-C\vert\langle\tilde S\rangle\langle\tilde{\bar N}\rangle\vert \\
-A\vert\langle\tilde{\bar S}\rangle\langle\tilde{S}\rangle\vert& 
A\vert\langle\tilde{\bar S}\rangle\vert^2 &
C\vert\langle\tilde{\bar S}\rangle\langle\tilde{\bar N}\rangle\vert \\
-C\vert\langle\tilde S\rangle\langle\tilde{\bar N}\rangle\vert &
C\vert\langle\tilde{\bar S}\rangle\langle\tilde{\bar N}\rangle\vert &
B\vert\langle\tilde{\bar N}\rangle\vert^2 \\
\end{array}\right), 
\label{eqhhh}
\end{equation}
where $A$, $B$ and $C$ are defined as
\begin{eqnarray}
&&A=(g_X^2Q_X^{S2}+g_A^2Q_A^{S2}), \qquad B=(g_X^2Q_X^{N2}+g_A^2Q_A^{N2}),
\nonumber \\
&&C=(g_X^2Q_X^{S}Q_X^{N}+g_A^2Q_A^{S}Q_A^{N}). 
\end{eqnarray}
The matrix ${\cal M}^2$ is written by the basis of 
$(S,~\bar S,~\bar N)$. 
The masses of spinor components of $(S,~\bar S,~\bar N)$ are generated through
the mixings with gauginos of $U(1)_X$ and U(1)$_A$.
Since $m_{\tilde{\bar N}}^2$ in eq.~(\ref{eqhhh}) 
is sufficiently large, the value of 
$\langle\tilde{\bar N}\rangle$ is
kept in the same one even after the introduction of 
the soft supersymmetry breaking effects.
On the other hand, mass eigenstates of ${\cal M}^2$ have 
an interesting feature.
One of them is massless and it is composed of $S$ and $\bar S$ 
such as $\Phi_0\propto \langle
\tilde{\bar S}\rangle S+ \langle\tilde S\rangle \bar S$.
The fact that each value of $\langle\tilde S\rangle$ and $\langle\tilde{\bar
S}\rangle$ is not determined in eq.~(\ref{eqh}) comes from the
existence of this mode.
The other mass eigenstates  $\Phi_1$ and $\Phi_2$ have 
masses of $O(\sqrt{\xi_A/F})$
and are composed as the linear combinations of $\langle
\tilde{\bar S}\rangle S- \langle\tilde S\rangle \bar S$ and $\bar N$.
A mass of the massless state $\Phi_0$ is considered to be supplied 
through the soft supersymmetry breaking effects and 
also the radiative correction on it, which are considered to be 
of the order of the weak scale. 
When its mass is generated, the scalar component of $\tilde\Phi_0$ 
starts to oscillate around the true vacuum keeping the condition for 
$\delta\tilde S_f^2$ in eq.~(\ref{eqh}) and its decay is expected to
induce the reheating. Since the mass of $\Phi_0$ is of the order of 
the weak scale, it is also expected to play a certain role in the 
particle physics phenomenology at a low energy scale.

By now we have briefly sketched the inflation scenario in this model.
We discuss more detailed quantitative features here.
The slow-roll parameters of the inflation defined as $\epsilon={1\over 2}
\tilde M_{\rm pl}^2 \left(V^\prime/V\right)^2$ and 
$\eta= \tilde M_{\rm pl}^2V^{\prime\prime}/V$ should satisfy the
conditions $\epsilon,~\vert\eta\vert~{^<_\sim}~1$ during 
the slow-roll period\footnote{
$\tilde M_{\rm pl}$ is a reduced Planck mass defined as $\tilde M_{\rm
pl}=M_{\rm pl}/\sqrt{8\pi}$.}.
The values of these slow-roll parameters at 
$\vert\tilde\phi\vert=\vert\tilde\phi_c\vert$ 
is estimated by using $V(\tilde\phi)$ in eq.~(\ref{eqgg}) as
\begin{equation}
\epsilon={k^2g_A^2\over 128\pi^4}{\tilde M_{\rm pl}^2 \over\xi_A}GF^3,
\qquad
\eta=-{k^2\over 8\pi^2}{\tilde M_{\rm pl}^2\over\xi_A}F.
\label{eqhh}
\end{equation}
The end of the slow-roll inflation is determined as the earlier period
of the realization of $\vert\tilde\phi\vert\simeq \vert\tilde\phi_c\vert$ or 
$\vert\eta\vert\simeq 1$. If we write the value of $\tilde\phi$ at that time
as $\tilde\phi_{\rm end}$, we find that 
$\vert\tilde\phi_c\vert\simeq\vert\tilde\phi_{\rm end}\vert$
is realized when $k^2 F \simeq8\pi^2\xi_A/\tilde M_{\rm pl}^2$ 
is satisfied.
The e-folds during the slow-roll inflation is written by using the
slow-roll approximation as
\begin{equation}
N(\tilde\phi)=\int^{\tilde\phi}_{\tilde\phi_{\rm end}}
{1\over\tilde M_{\rm pl}^2}{V\over V^\prime}~ d\tilde\phi
={4\pi^2\over g_A^2}{1\over GF^2}{1\over \tilde M_{\rm pl}^2}
\left(\vert\tilde\phi\vert^2-\vert\tilde\phi_{\rm end}\vert^2\right),
\label{eqhi}
\end{equation}
and then $\vert\tilde\phi\vert^2\simeq { 1 \over
4\pi^2}N(\tilde\phi)g_A^2GF^2\tilde M_{\rm pl}^2$ where we used the relation 
$\displaystyle\vert\tilde\phi\vert
\gg \vert\tilde\phi_{\rm end}\vert$. In eq.~(\ref{eqhi}),
$\tilde\phi$ is a value at the period when a relevant cosmological scale
of the universe leaves the horizon.\footnote{
Unless $g_A^2GF^2$ is sufficiently smaller than 1, we cannot have
sufficient e-folds such as $N\sim 50$ keeping a condition
$\vert\tilde\phi\vert \ll \tilde M_{\rm pl}$, which guarantees the
validity to treat the system by the field theory. 
However, as we will see next, it cannot be small enough. } 

The CMB data requires that $\xi_A$ should satisfy 
a suitable condition.
The COBE normalization for the density perturbations is given as
$\delta_H=\sqrt{V/(150\pi^2\tilde M_{\rm pl}^4\epsilon)}
\sim 1.95\times 10^{-5}$ \cite{cobe}, where $V$ and $V^\prime$ are 
also estimated at the epoch where the COBE scale leaves the horizon.
By using eq.~(\ref{eqhi}), we find that
this imposes the condition\footnote{We ignore
any gravitational wave contributions since it can be safely neglected 
in the present model.} 
\begin{equation}
\xi_A\simeq 8.5\times 10^{-6}\left({50\over N(\tilde\phi)}
\right)^{1\over 2} F \tilde M_{\rm pl}^2.
\label{eqi}
\end{equation}
Combining a result given below eq.~(\ref{eqhh}) and eq.~(\ref{eqi})
which is the result of the COBE normalization, we find that 
$\vert\tilde\phi_c\vert\simeq \vert\tilde\phi_{\rm end}\vert$ is
realized for $k$ such as $k_0\equiv 
0.025\left({50\over N}\right)^{1/2}$.
If $k$ is smaller than this value, $\vert\tilde\phi_c\vert >
\vert\tilde\phi_{\rm end}\vert$ is satisfied and the fast-roll
inflation period appears additionally after the end of the slow-roll
period. To make the situation simple, we assume 
$\displaystyle k\simeq k_0$ in the
following part as far as we do not mention it.
 
As the usual hybrid inflation and $D$-term inflation, 
this model predicts an almost flat spectrum of the density perturbations.
In fact, by using eqs.~(\ref{eqhi}) and (\ref{eqi}) the spectral index is 
estimated as \cite{index}
\begin{equation}
n-1=2\eta-6\epsilon\simeq -{1\over N(\tilde\phi)}
\left(1+{3\over 16\pi^2}g_A^2GF^2\right).
\label{eqk}
\end{equation}
This shows a good agreement with the analysis of the recent observations
\cite{spectral}.
Although the second term in the parentheses of eq.~(\ref{eqk}) 
might be rather large and produce a non-negligible effect
depending on the value of $F$,
it is natural to consider that $g_A^2GF^2$ is at most $O(1)$.
Then this term is expected to be small enough.
The deviation of the spectral index $n$ from 1 is estimated 
as $-1/N(\tilde\phi)$, 
which is the well-known feature of the $D$-term inflation \cite{index}.
This feature may be useful to discriminate this model 
from others also.

If $ F =O(1)$ is assumed to be satisfied, 
for example, eq.~(\ref{eqi}) shows that $\xi_A$ takes a value 
of order $10^{16}$~GeV.  
On the other hand, if the F-I $D$-term $\xi_A$ is assumed to have its origin 
in the anomalous U(1) in the weakly-coupled superstring, 
$\xi_A$ can be written as \cite{xi}
\begin{equation}
\xi_A={{\rm Tr}~Q_A \over 192\pi^2}g_{\rm st}^2\tilde M_{\rm pl}^2.
\label{eqj}
\end{equation}
It takes a value in the range $10^{17-18}$~GeV for $g_{\rm st}\sim 0.1
- 1$ and ${\rm Tr}~Q_A\sim O(100)$, which are usually considered to be 
the typical values. In the usual $D$-term inflation model, 
there is a discrepancy between this and the above mentioned value 
required by the CMB data \cite{discrep}.
In this model, however, the effective $\xi_A$ which dominates 
the vacuum energy contributing to the density perturbations 
has an additional factor $F$ as seen in eq.~(\ref{eqi}).
Because of this factor, as suggested in \cite{err}, there may be a 
potential possibility to reconcile this discrepancy\footnote{
We do not consider a kinetic term mixing between U(1)$_X$ and U(1)$_A$
\cite{kinetic}. If we take it into account, the situation for this discrepancy
seems to be worse.} as far as $ F$ takes a value 
$F\sim 6\times 10^3\left({N(\tilde\phi)/ 50}\right)^{1/2}
g_{\rm st}^2$.
A rather large $F$ seems to be obtainable if $Q_A^S \gg Q_X^S$ and 
$g_X \simeq g_A(=g_{\rm st})$ is satisfied.
In such a case, both $G\ll 1$ and $F \gg 1$ can be satisfied as far
as $Q_X^N>O(1)$, as can be easily seen from their expressions 
in eqs.~(\ref{eqd}) and (\ref{eqg}).
Under this situation, $g_A^2GF^2$ can be estimated as 
$\sim g_X^2Q_X^{N2}$ and it cannot take a small value so as 
to make $\vert\tilde\phi\vert$ sufficiently small in comparison with $\tilde 
M_{\rm pl}$. We must consider the case with
$\vert\tilde\phi\vert~{^<_\sim}~\tilde M_{\rm pl}$ and
the validity to treat the system by the field theory seems to be
marginal. It depends on the value of $g_{\rm st}$.
If the Yukawa coupling $k$ in $W_1$ is smaller than $k_0$, however, 
there is an additional fast-roll inflation after this slow-roll
inflation. In this case $N(\tilde\phi)$ can be smaller and as a result
 $F$ can be somewhat smaller. Thus the smaller values 
of $k$ and $g_{\rm st}$ seems to give a promising possibility 
for the reconciliation or the relaxation of 
the discrepancy between the required $\xi_A$ values . In such a situation,
$\vert\tilde\phi\vert<\tilde M_{\rm pl}$ is satisfied and $\vert n-1\vert$
becomes somewhat larger because of the smaller value of $N(\tilde\phi)$
required for the slow-roll inflation.

Next we consider the reheating in this model. As mentioned in the
previous part, $\tilde\phi$ and the scalar components of two massive states 
$\Phi_1,~\Phi_2$ start to oscillate around the global minimum 
when $\vert\tilde\phi\vert~{^<_\sim}~\vert\tilde{\phi_c}\vert$ is realized.
Preheating and reheating processes are expected to proceed
through these oscillations. 
On the preheating related to $\tilde\phi$ and $\tilde{\bar N}$ 
in this model, the study in \cite{bl} is very useful.
The hybrid inflation model with the scalar potential
\begin{equation}
V(\phi,\sigma)={1\over 4\lambda}(\lambda\sigma^2-M^2)^2
+{1\over 2}g^2\phi^2\sigma^2+{1\over 2}m^2\phi^2 
\end{equation}
has a very similar structure to the scalar sector in the present model.
If we use the following replacement in the scalar potential in \cite{bl}: 
\begin{equation}
\phi \rightarrow\tilde\phi, 
\quad \sigma\rightarrow\tilde{\bar N}, \quad
M^2 \rightarrow -g_A^2Q_A^N\xi_A, \quad
g^2\rightarrow k^2, \quad  \lambda\rightarrow g_X^2Q_X^{N2}+g_A^2Q_A^{N2},
\end{equation} 
a part of the scalar potential (\ref{eqb}) can be obtained. 
We find that our model corresponds to a case of $\lambda \gg g^2$ 
in \cite{bl} for this sector by taking account of the previous
discussion on the couplings.
Following their study,
the oscillation energy is expected to be concentrated immediately 
to the oscillation of $\phi$ in that case. 
They also shows that there is no effective 
preheating and no significant particle production of $\phi$ and $\sigma$
through its oscillation. We expect that their results can be
applied to our model and the preheating effect can be also neglected here.  
Then the deviation of $\tilde\Phi_1$ and $\tilde\Phi_2$ which are
composed of $\tilde S,~\tilde{\bar S}$ and $\tilde{\bar N}$ from their
true vacuum values during the inflation cannot cause the oscillation 
since their oscillation energies are expected to be instantaneously 
transferred into the oscillation of
$\tilde\phi$ through the coupling with $\tilde{\bar N}$.
We need to consider only the reheating due to the usual perturbative 
$\tilde\phi$ decay.  
The decays of the inflaton $\tilde\phi$ into
$N$ and $\Phi_{1,2}$ is expected to be kinematically prohibited 
in the present model since 
$m_{\Phi_1}^2,~m_{\Phi_2}^2 > m_{\tilde\phi}^2$ is satisfied 
following the discussions below eqs.~(\ref{eqhhh}) and (\ref{eqj}).
Moreover, since $\Phi_0$ has no $\bar N$ component in it, 
$\tilde\phi$ cannot decay into $\Phi_0$ through the coupling 
$k\phi\bar NN$.
Thus we cannot expect the reheating by the $\tilde\phi$-decay 
into the light MSSM particles through the couplings in the 
renormalizable superpotential. 
However, in general, its oscillation can decay into the MSSM light fields 
through the allowed nonrenormalizable couplings contained in the 
superpotential which may be expressed as 
\begin{equation}
W_2=c~{\phi\psi^r\over\tilde M_{\rm pl}^{r-2}}
\end{equation}
where $\psi$ stands for the MSSM chiral superfields and $c$ is
assumed to be $O(1)$.
If $\tilde\phi$ has a decay width $\Gamma_{\tilde\phi}$ through this
type of coupling, the reheating temperature can be
estimated by assuming the Hubble parameter $H$ to be $H\sim
\Gamma_{\tilde\phi}$ as \cite{kr} 
\begin{equation}
T_{\rm RH}^{\tilde\phi}\sim {c~ k^{r-{3\over 2}}\over g_{_\ast}^{1/4}
 (3.6\times 10^2)^{r-4}}
\left({P_r\over P_4}\right)^{1\over 2}\times 1.3\times 10^{10}~{\rm GeV},
\label{eqx} 
\end{equation}
where $g_\ast$ is an effective number of the relativistic degrees of
freedom at this period and $P_r/P_4$ represents the ratio of the 
phase space factors
between $r$- and 4-body decays, which decreases with the increase of $r$.
As far as $r \ge 4$ is satisfied, the constraint coming from the 
gravitino problem can be satisfied. The decay of 
$\tilde\phi$ occurs at $H~{^<_\sim}~M_{\rm susy}$ depending on the value of
$r~(\ge 4)$, where $M_{\rm susy}$ stands for a typical soft
supersymmetry breaking scale of order 1 TeV.
The value of $r$ and the couplings of $S$ and $\bar S$ to the light MSSM
fields determine whether the condensate of $\tilde\phi$ can decay 
earlier than the deviation of $\Phi_0$ from the true vacuum or not. 
This largely affects the baryogenesis scenario in this model. 
For this study we need to fix
the couplings of $S$ and $\bar S$ with the MSSM fields in the 
nonrenormalizable terms of the superpotential.

\section{Implications for particle physics}
We need to introduce additional interactions of the SM singlet chiral 
superfields ($N$, $\bar N$) and ($S$, $\bar S$)
with the MSSM chiral superfields in the superpotential 
without modifying the features of the inflation
in order to fix their role in the particle physics phenomenology. 
Taking account that both $\tilde S$ and $\tilde{\bar S}$ can have 
nonzero vacuum expectation values (VEVs) at the
true vacuum as shown in eq.~(\ref{eqh}), we consider the following 
terms as the dominant 
ones in the superpotential,
\begin{equation}
W_3=h{(S\bar S)^\ell\over \tilde M_{\rm pl}^{2\ell}}L_\alpha H_2\bar N 
+ \lambda {(S\bar S)^n\over \tilde M_{\rm pl}^{2n}}SH_1H_2
+ \gamma {(S\bar S)^m\over \tilde M_{\rm pl}^{2m-3}},
\label{eql}
\end{equation}
where the powers $\ell,~m$ and $n$ are assumed 
to satisfy $\ell,~n\ge 0$ and $m \ge 2$, respectively.
The coupling constants $h$, $\lambda$ and $\gamma$ should be  
assumed to be $O(1)$ at least when the couplings are 
nonrenormalizable.\footnote{We do 
not consider the generation dependence of $h$, for simplicity.
The third term is introduced to stabilize the potential
for $\tilde S$ and $\tilde{\bar S}$. It could contribute the
inflation potential (\ref{eqb}). However, we can check its effect is not 
sufficiently large to change the present inflation scenario.}
A chiral superfield $L_\alpha$ stands for a doublet lepton and $H_1,~H_2$ are 
the usual doublet Higgs chiral superfields.
The addition of these terms does not affect the above discussed 
inflation scenario. 
It is useful to note that the usual $D$-term flatness along the
well-known directions $H_1H_2$ and $LH_2$ in the MSSM is lost 
because of the additional U(1) interactions. 
Thus, differently from the MSSM, the Affleck-Dine leptogenesis \cite{ad1}
along the latter direction cannot be expected in this model.
The model is naturally expected to have the soft supersymmetry 
breaking parameters corresponding to the superpotential $W_3$ such as
\begin{equation}
{\cal L}_{\rm soft}= A_h{(\tilde S\tilde{\bar S})^\ell\over 
\tilde M_{\rm pl}^{2\ell}} 
\tilde L_\alpha \tilde H_2\tilde{\bar N} 
+A_\lambda{(\tilde S\tilde{\bar S})^n\over 
\tilde M_{\rm pl}^{2n}} 
\tilde S \tilde H_1\tilde H_2 +\cdots,
\end{equation}
where $A_h$ and $A_\lambda$ are assumed to be $O(M_{\rm susy})$ and complex.

\subsection{$\mu$-term and neutrino masses}
The structure of the superpotential $W_3$ crucially affects the
phenomenology at a low energy region.
For example, the second term in $W_3$ may be considered to induce
an effective $\mu$-term of the MSSM in the form as 
$\mu={\lambda \over \tilde M_{\rm pl}^{2n}}
\langle\tilde S\rangle^{n+1}\langle\tilde{\bar S}\rangle^{n}$ 
if both $\tilde S$ and $\tilde{\bar S}$ get suitable VEVs. 
The ratio $\tan\alpha_S\equiv\langle\tilde S\rangle/\langle
\tilde{\bar S}\rangle$ 
determines the composition of $\Phi_0$ and also
a value of the reheating temperature $T_{\rm RH}^{\tilde S}$ 
due to the decay of the condensate of $\tilde \Phi_0$
as seen later.
As mentioned below eq.~(\ref{eqhhh}), there is a massless state
$\Phi_0$ which contains $S$ with the weight $\cos\alpha_S$.
The mass of $\tilde\Phi_0$ has the dominant contributions 
only from the soft supersymmetry breaking squared mass, 
radiative corrections based on the supersymmetry breaking effects and
also the last term in $W_3$.
If its squared mass becomes negative 
by the radiative correction through the couplings with extra colored 
chiral superfields $g$ and $\bar g$ such as $Sg\bar g$ in 
the superpotential \cite{sy2,rad1} or by the suitable supersymmetry breaking
\cite{zs}, for example, the values of 
$\langle\tilde S\rangle$ and $\langle\tilde{\bar S}\rangle$
are expected to be 
shifted into the true vacuum ones satisfying the relation 
for $\delta \tilde S_f^2$ in eq.~(\ref{eqh}).
Thus, as far as we do not consider the large cancellation between the
contributions from $\langle\tilde S\rangle$ and 
$\langle\tilde{\bar S}\rangle$, either $\langle\tilde S\rangle$ or 
$\langle\tilde{\bar S}\rangle$ are at least required to be an
almost GUT scale.\footnote{This symmetry breaking scale is somewhat 
larger than the one assumed in the similar model where the 
$\mu$-problem and the strong $CP$ problem 
have been considered \cite{gk}-\cite{dls}.}
The value of $\tan\alpha_S$ is expected to take a rather wide range
value \cite{rad1} depending on the details of the model, 
for example, the extra matter contents coupled with $S$ and $\bar S$ 
and the power $m$, and also the nature of the symmetry breaking 
in this sector.
There seem to be two possibilities to determine the values of
$\langle \tilde S\rangle$ and $\langle \tilde{\bar S}\rangle$.
One is the case where the third term in $W_3$ plays a crucial role
in their determination. The other case is based on the pure radiative 
effects. In that case the value of the power $m$ is large 
enough and then the third term 
in $W_3$ can have only the much smaller effect than the radiative correction 
\cite{rad1}.
Although the latter possibility needs the numerical study 
to get some results,\footnote{
This study is beyond the scope of the paper and 
we do not discuss this case further.} 
the former one allows us to have a rough estimation of the symmetry
breaking scale. 

For this instruction, let us present an example which is found to 
be interesting from some phenomenological reasons as shown later.
We consider the situation such that $\Phi_0$ is dominated by $S$ and then
$\langle \tilde S\rangle \ll \langle \tilde{\bar S}\rangle\sim
O(10^{15})$~GeV is satisfied. We also require the realization of a 
suitable $\mu$ value such as $O(1)$~TeV.
Then, if we assume that the third term in $W_3$ can fix the 
values of $\langle \tilde S\rangle$ and $\langle \tilde{\bar S}\rangle$,
we find that $\tan\alpha_S=\lambda^{-1\over n+1}10^{6(n-2)\over n+1}$ 
should be satisfied and the allowed value of $n$ is restricted 
to zero or one. As seen later, the phenomenological condition for the 
reheating temperature can give a severe constraint on the value of $n$. 
Here let us assume the masses of $\tilde S$ and
$\tilde{\bar S}$ to be $m_{\tilde S}^2<0$ and $m_{\tilde{\bar S}}^2<0$ due
to some radiative effects. 
The potential minimization gives 
\begin{eqnarray}
&&\vert\langle \tilde S\rangle\vert \sim \tilde M_{\rm pl}^{2m-3\over 2(m-1)}
\vert m_{\tilde S}\vert^{-1 \over 2}
\vert m_{\tilde{\bar S}}\vert^{m\over 2(m-1)}, \nonumber \\
&&\vert\langle \tilde{\bar S}\rangle\vert 
\sim \tilde M_{\rm pl}^{2m-3\over 2(m-1)}
\vert m_{\tilde S}\vert^{ 1\over 2}
\vert m_{\tilde{\bar S}}\vert^{2-m\over 2(m-1)}.
\end{eqnarray}
From this result, we find $\tan\alpha_S\sim \vert m_{\tilde{\bar S}}\vert/
\vert m_{\tilde S}\vert$. The situation with 
the suitable value of $\tan\alpha_S$ is realized in the case of $m=3$ for the 
appropriately tuned scalar masses in the case of $n=1$.\footnote{In the
case of $n=0$, we have $\tan\alpha_S=10^{-12}$ and its realization
requires extremely tuned $m_{\tilde S}^2$ and $m_{\tilde{\bar S}}^2$. } 
The situation may be somewhat different from this 
in the pure radiative symmetry breaking case.
 
In the model defined by $W_1+W_2+W_3$ the singlet field $\bar N$ can be 
understood as a charge conjugated field of a right-handed neutrino and it 
has a lepton number $-1$. Thus
we can consider a neutrino mass generation based on the first 
coupling in $W_3$. 
Since we have a large Majorana mass for the right-handed neutrino 
$\bar N$ as shown in eq.~(\ref{eqhhh}), 
the ordinary seesaw mechanism \cite{seesaw} might be expected to work.
In fact, $\bar N$ gets a large Majorana mass of
$O(\vert\langle\tilde{\bar N}\rangle\vert)$ and then we can obtain a 
small neutrino mass, which is estimated for the $\ell=0$ case, as
\begin{equation}
m_\nu\sim {h^2\vert\langle\tilde H_2\rangle\vert^2\over 
\vert\langle\tilde{\bar N}\rangle\vert}. 
\label{eqp}
\end{equation}
If we impose $m_\nu\sim O(10^{-1})$~eV based on the atmospheric neutrino
analysis, we find that $h$ should satisfy 
$h\sim 0.01$ by using eq.~(\ref{eqh}).
However, the same term in eq.~(\ref{eql})
spontaneously induces a bilinear $R$-parity violating term 
$\varepsilon LH_2$ with $\varepsilon=h\langle\tilde{\bar N}\rangle$. 
If we take $h\sim 0.01$, $\varepsilon$ becomes too large and 
such a large $R$-parity violating term is forbidden by various
phenomenological constraints \cite{rbreak}.
This means that the magnitude of the Yukawa coupling constant $h$  
required by the phenomenological condition coming from the 
$R$-parity violation largely contradicts with
the one required by the neutrino mass.

Fortunately, we can consider another possibility for the generation of 
neutrino masses. Neutrino masses may be produced by
neutrino-neutralino mixings as discussed in \cite{rparity1}-\cite{rparity3}.
If $\varepsilon$ is sufficiently small as a result of the higher
orderness of the relevant term in the superpotential $W_3$, 
a weak scale seesaw mechanism is expected to induce neutrino masses as 
\begin{equation}
m_\nu \simeq {\varepsilon^2 \over \mu }
\simeq (\tan\alpha_S)^{2\ell-n-1}\left({ \vert\langle\tilde{\bar S}
\rangle\vert 
\over \tilde M_{\rm pl}}\right)^{4\ell-2n} 
{\vert\langle\tilde{\bar N}\rangle\vert^2\over 
\vert\langle\tilde{\bar S}\rangle\vert},
\label{eqq}
\end{equation}
where we use that $h$ and $\lambda$ are $O(1)$. Since 
$\vert\langle\tilde{\bar S}\rangle\vert\sim
\langle\vert\tilde{\bar N}\rangle\vert\sim
O(10^{15})$~GeV should be satisfied in this inflation model, we find
that $m_\nu$ can be in the suitable range of 
$O(10^{-1})$~eV for the atmospheric neutrino observation 
by taking $\ell=2$ and $n=1$. In this case  
$\vert\varepsilon\vert\sim 10^{-6}\vert\mu\vert$ 
and $\mu=O(1)$~TeV are
satisfied. \footnote{Here we give a naive estimation of $\varepsilon$,
as an example. However, as discussed in \cite{rparity4}, the
neutrino mass cannot be expressed in such a simple form as
$\varepsilon^2/\mu$ in the present model. 
Following their detailed analysis, we can have
$O(10^{-1})$~eV neutrino mass even for $\vert\varepsilon\vert\sim
10^{-4\sim -3}\vert\mu\vert$.}
Moreover, this ratio of $\varepsilon$ 
and $\mu$ is independent of the value of $\langle\tilde S\rangle$ 
in this case, which is useful to note for the later discussion.  
Since the extreme smallness of the effective Yukawa coupling of the
neutrino due to the nonrenormalizablity makes the 
above mentioned ordinary seesaw mass (\ref{eqp}) is negligible, 
the neutrino mass and mixing should be considered on the basis of 
the neutrino-neutralino mixing along the discussions in 
\cite{rparity1}-\cite{rparity3}.
It is interesting that in this scenario the $\bar N$ sector 
is not necessarily required
to be extended into the multi-generation in order to explain the 
required pattern of masses and mixings of neutrinos. 
Thus we can directly apply the present inflation scenario 
to the explanation of small neutrino masses.

In this case it is expected the appearance of a physical Nambu-Goldstone 
boson after the singlet fields $\tilde{\bar N}$, $\tilde S$ and 
$\tilde{\bar S}$ get the VEVs. Its physical feature is completely
dependent on its composition. If it is dominantly composed of
$\tilde{\bar N}$, it may behave like a Majoron. However, it seems to be hard 
to find its physical implication since its effective couplings to the ordinary
matter fields are too weak. On the other hand, if it is dominated by
$\tilde S$ or $\tilde{\bar S}$, it may behave like an axion related to
the strong CP problem. In such a case its consistency with various
constraints depends on the details of the model and its study seems to
be beyond the scope of this work. We will discuss this point another place. 

If we exchange the role of $\tilde N$ and $\tilde{\bar N}$ in eq.(\ref{eqh}) 
by modifying their charge assignments for U(1)$_X\times$ U(1)$_A$,
there appears a new possibility for the generation of neutrino masses. 
In this case $\tilde{\bar N}$ has no VEV at the global minimum 
and then we have no bilinear $R$-parity violating term.
However, its fermion partner gets a large Majorana mass 
$k\vert\langle\tilde N\rangle\vert$ through the mixing with $\phi$ in $W_1$.
Small neutrino masses can be explained by the ordinary seesaw 
mechanism in the case of $\ell=0$ which corresponds to the renormalizable 
neutrino Yukawa coupling.
They can be estimated as
\begin{equation}
m_\nu\sim {h^2\vert\langle\tilde H_2\rangle\vert^2\over 
k\vert\langle\tilde N\rangle\vert}. 
\label{eqs}
\end{equation}
If we impose $m_\nu\sim O(10^{-1})$~eV, $h$ should satisfy 
$h\sim10^{-2}k^{1/2}$.
In this case, however, unless we introduce additional $\bar N$'s, we need 
some new mechanism to explain the required pattern of 
masses and mixings of neutrinos. 
On the other hand, if we introduce new $\bar N$'s, some modification
will be required in the above inflation scenario. 
Because of these reasons the neutrino mass generation based on the
neutrino-neutralino mixing seems to be the most promising
in the present inflation scenario.

We find that the superpotential $W_3$ with $\ell=2$ and $n=1$ is
promising from the $\mu$-problem and neutrino masses in the present scenario.
This kind of favorable structure of the superpotential $W_1+W_2+W_3$ is 
reproduced consistently with the Yukawa couplings in the MSSM
by introducing a suitable discrete symmetry. 
For example, as a simple candidate for such a discrete symmetry, 
we can consider the case in which the MSSM matter fields have the same charge.
We take the symmetry as $Z_{24}$ and adopt the following charge
assignment:\footnote{There are many other possibilities. We only present
this example to show that the control of the superpotential by the
discrete symmetry is not so difficult in this model.}
\begin{eqnarray}
&&Q_Z(Q_\alpha,~L_\alpha,~\bar U_\alpha,~\bar D_\alpha,~
\bar E_\alpha)=3, \qquad Q_Z(H_1,~H_2)=-6,\nonumber \\
&&Q_Z(N)=1, \quad Q_Z(\bar N)=-13, \quad  Q_Z(S,~\bar S)=4, 
\quad  Q_Z(\phi)=12,
\label{eqw}
\end{eqnarray}
where $\alpha$ is a generation index.
This discrete symmetry realizes the superpotential $W_1$ and $W_3$ with
$\ell=2,~n=1,~m=3$ and also $W_2$ with $r=4$.
 
\subsection{Generation of baryon number asymmetry}
Finally we discuss the baryogenesis in this model.
After the inflation, the condensate of $\tilde{\bar N}$ 
(or $\tilde\Phi_{1,2}$) immediately reduces to a true
vacuum value and it cannot effectively oscillate as mentioned before. 
As a result, we cannot use its decay for the leptogenesis. 
On the other hand, since a scalar component of the massless 
state $\Phi_0$ is deviated from its true vacuum value after its 
squared mass becomes negative, $\tilde\Phi_0$ starts to oscillate 
around its true vacuum value at $H\sim m_{\tilde\Phi_0}$.
The leptogenesis may be expected to occur through the decay of
$\tilde\Phi_0$ accompanied by its oscillation. 
The oscillation of $\tilde\Phi_0$ which contains $\tilde S$ and
$\tilde{\bar S}$ decays into 
$L$ and $H_2$ or $H_1$ and $H_2$ 
through the first or second couplings in eq.~(\ref{eql}). 
Following this decay, $L$, $H_1$ and $H_2$ are also transfered 
into the other light SM fields and they also contribute to the 
thermalization of the universe.
The decay width of $\tilde S$ is estimated as 
$\Gamma_{\tilde S}\sim {1\over 4\pi}f^2M_{\rm susy}$
because of $m_{\tilde\Phi_0}\sim M_{\rm susy}$.
The effective coupling $f$ for the ${\tilde S}$ decay into each mode 
should be understood as
$h_{\rm eff}={h\over \tilde M_{\rm pl}^{2\ell}}
 (\tan\alpha_S)^{\ell-1}\langle\tilde{\bar S}\rangle^{2\ell-1}
\langle\tilde {\bar N}\rangle$ or
$\lambda_{\rm eff}={\lambda\over \tilde M_{\rm pl}^{2n}}
(\tan\alpha_S)^n\langle\tilde{\bar S}\rangle^{2n}$, respectively.
Depending on the relative strength of these effective couplings 
$h_{\rm eff}$ and $\lambda_{\rm eff}$, the main decay mode of the
$\tilde\Phi_0$ condensate is determined.
Since we found that $W_3$ with $\ell=2$ and $n=1$ 
is favorable for the explanation of both the $\mu$-problem and 
the small neutrino masses in the previous discussion,
we will concentrate our attention on this possibility.
In this case the $\tilde\Phi_0$ condensate mainly composed of $\tilde S$ 
 decays into 
$H_1$ and $H_2$ through the second effective couplings 
$\lambda_{\rm eff}\tilde S H_1 H_2$.
In this decay the reheating temperature can be estimated as 
\begin{equation}
T_{\rm RH}^{\tilde S}\simeq {1.7 \over g_{_\ast}^{1/ 4}}
\sqrt{\Gamma_{\tilde S}M_{\rm pl}}
\sim{0.48 \over g_{_\ast}^{1/ 4}}\lambda_{\rm eff}
\sqrt{\tilde M_{\rm pl} M_{\rm susy}}.
\end{equation}
This reheating temperature should satisfy a certain upper bound to prohibit 
the over production of gravitinos \cite{grav} and also it is high enough 
to allow us to use the 
electroweak sphaleron interaction for the production of 
the baryon number asymmetry.
These requirement give a condition on $T_{\rm RH}^{\tilde S}$ such as 
$10^2~{\rm GeV}~{^<_\sim}~T_{\rm RH}^{\tilde S}~{^<_\sim}~10^9~{\rm
GeV}$.
This means that $\lambda_{\rm eff}$ has to satisfy 
$10^{-9}~{^<_\sim}~\lambda_{\rm eff}~{^<_\sim}~10^{-2}$.
This condition can be satisfied in the above discussed $n=1$ case 
where the nonrenormalizable term determines the vacuum.
In fact, since we have 
$\vert\langle \tilde{\bar S}\rangle\vert=O(10^{15})$~GeV and
$\tan\alpha_S=10^{-3}$ in this case, 
$\lambda_{\rm eff}=O(10^{-9})$ and 
$T_{\rm RH}^{\tilde S}\sim O(10^2)$~GeV is realized for $g_\ast=O(10^2)$.
The pure radiative symmetry breaking case may be also expected to 
be able to satisfy this condition similarly. 

We may be possible to consider several baryogenesis scenarios through the 
electroweak sphaleron interaction. The first possibility is the usual
electroweak baryogenesis scenario in the electroweak phase transition.
The $\mu$ term is effectively generated through the coupling with the SM 
singlet fields, the phase transition can have the stronger first
orderness in comparison with the MSSM case because of the existence of the
trilinear coupling of the Higgs scalar fields as in the case of the NMSSM.
This possibility will, however, be expected only in the case where either
$\langle\tilde S \rangle$ or $\langle\tilde{\bar S}\rangle$ takes a
value not so far from the weak scale. 

The second possibility may be considered on the basis of the decay 
of the oscillation of $\tilde\Phi_0$.
In this case the problem is whether the decay product can bring the 
baryon number asymmetry or not. 
Since the oscillation of $\tilde\Phi_0$ is non-thermal, this decay
proceeds in the way of the out-of-thermal equilibrium. 
Thus, if there are some $CP$ violating complex phases and
$\Gamma_{\tilde S}<H_{T=m_{\tilde\Phi_0}}$ is satisfied,
we can expect the production of the asymmetry between the Higgsino 
$H_{1,2}$ number density and the anti-Higgsino 
$H_{1,2}^c$ number density through the decay by the
second term in eq.~(\ref{eql}).
This out-of-equilibrium condition restricts the region of $\lambda_{\rm
eff}$ obtained from the condition for $T^{\tilde S}_{\rm RH}$ further
into 
\begin{equation}
10^{-9}~{^<_\sim}~\lambda_{\rm eff}~{^<_\sim}~10^{-7}.
\label{eqxyx}
\end{equation}
We should note that our concrete example satisfies this condition.
The pure radiative symmetry breaking may be able to loose this condition 
somewhat \cite{rad1}.
The Higgsino asymmetry $\epsilon_S$ can be 
produced dominantly by the interference
term between a tree diagram of $\tilde S\rightarrow H_1 H_2$ 
and a one-loop contribution coming from a diagram with an
$A_\lambda$ vertex and a gaugino $\lambda_{1,2}$ internal line. 
It can be roughly estimated as \cite{asym}
\begin{equation}
\epsilon_S\equiv{\Gamma_u(\tilde S\rightarrow H_1 H_2)
-\Gamma_u(\tilde S^c\rightarrow H_1^c H_2^c)\over 
\Gamma_u(\tilde S\rightarrow H_1 H_2)
+\Gamma_u(\tilde S^c\rightarrow H_1^c H_2^c)}
\sim
{g_2^2+g_1^2\over 16\pi}{\vert A_\lambda\vert\over m_{\tilde\Phi_0}}
\sin\delta,
\end{equation}
where $\delta={\rm arg}(A_\lambda)$.
Thus, by using the relations $\rho_{\rm tot}=sT_{\rm RH}$ 
and $\rho_{\tilde\Phi_0}=n_{\tilde\Phi_0}m_{\tilde\Phi_0}$,
the ratio of the total asymmetry of the $H_2$ number density 
to the entropy density can be estimated as
\begin{equation} 
{n_{\tilde H_2}-n_{\tilde H_2}^c\over s}
=\epsilon_S~{\rho_{\tilde\Phi_0}\over\rho_{\rm tot}}
{T_{\rm RH}^{\tilde\Phi_0}\over m_{\tilde\Phi_0}}
\simeq {g_2^2+g_1^2\over 16\pi}
{T_{\rm RH}^{\tilde\Phi_0}\over M_{\rm susy}}\sin\delta,
\label{eqxyz} 
\end{equation}
where $\rho_{\tilde\Phi_0}$ and $\rho_{\rm tot}$ are the
energy density stored in the $\tilde\Phi_0$ oscillation and 
the total energy density of the universe, respectively. 
The energy density of the universe is assumed to be occupied by the
oscillation energy of $\tilde\Phi_0$ at this period and then 
$\rho_{\rm tot}\sim \rho_{\tilde\Phi_0}$.

Here we should remind that we have a bilinear $R$-parity
violating term which comes from the first term in eq.~(\ref{eql}).
This term induces the mixing of $O(\vert\varepsilon\vert/\vert\mu\vert)$
between $L_\alpha$ and $H_2$.
If the lightest neutralinos and charginos are dominantly composed of the 
$H_2$ components, 
the produced Higgsino asymmetry $\epsilon_S$ is expected to be 
transferred into the lepton number asymmetry through this 
lepton number violating mixing $\varepsilon L_\alpha H_2$. 
We consider the case where these neutralino and chargino need longer
time to reach the thermal equilibrium through various interactions
in comparison with the typical sphaleron interaction time at 
this period.\footnote{This condition may be in a delicate position
in this model. For more precise analysis, we need to check it
numerically. 
The ambiguity related to this point will be taken into account as the
washout factor $\kappa$ here.}
Then there is sufficient time for 
the sphaleron interaction in the thermal equilibrium 
to convert the Higgsino $H_2$ asymmetry into the baryon asymmetry 
effectively through this lepton number violating mixing. 
By using this and eq.~(\ref{eqxyz}), the baryon number asymmetry 
generated through this process can be expected to be 
\begin{equation}
Y_B={n_B-n_{\bar B}\over s}\simeq c_s\kappa~{\vert\varepsilon\vert^2
\over\vert \mu\vert^2} 
\left({g_2^2+g_1^2\over 16\pi} 
{T_{\rm RH}^{\tilde\Phi_0}\over M_{\rm susy}}\sin\delta\right)
\sim 3\times 10^{-4\sim -2}\lambda_{\rm eff}\kappa\sin\delta,
\label{eqyy}
\end{equation}
where $\kappa$ is the washout factor and $c_s$ is the conversion 
rate of the lepton number asymmetry into
the baryon number asymmetry by the sphaleron interaction and takes a
value of $c_s=O(1)$ \cite{lb}. To derive a last numerical factor, we used
the value $\vert\varepsilon\vert/\vert\mu\vert=10^{-4\sim -3}$ 
which has been referred in the footnote below eq.~(\ref{eqq}). 
If there is no entropy production after the decay of $\tilde\Phi_0$
condensate, this baryon number asymmetry is understood to represent 
the one of the present universe. 
Comparing this with the present predicted value 
$Y_B=(0.6~-~1))\times 10^{-10}$, we find that 
$2\times 10^{-9}~{^<_\sim}~
\lambda_{\rm eff}\kappa\sin\delta ~{^<_\sim}~ 3\times 10^{-7}$ should be 
satisfied.
By substituting this into eq.~(\ref{eqxyx}), we obtain 
$2\times 10^{-2}~{^<_\sim}~\kappa\sin\delta~{^<_\sim}~1$.
It is noticeable that this condition can be satisfied by the reasonable
values of $\sin\delta$ and $\kappa$.

As we noted at the end of last section, however, there can be an additional 
reheating due to the decay of the $\tilde\phi$ condensate into the 
light SM fields. If this decay occurs after the decay of the 
$\tilde\Phi_0$ condensates, the entropy produced by 
this $\tilde \phi$ decay will dilute the baryon number asymmetry obtained 
from eq.~({\ref{eqyy}}) in such a way as \cite{kr}
\begin{equation}
Y_B\sim {g_2^2+g_1^2\over 16\pi}c_s\kappa
{\vert\varepsilon\vert^2 \over\vert\mu\vert^2}\sin\delta
{T_{\rm RH}^{\tilde\phi}\over M_{\rm susy}}
\sim 3\times 10^{-11\sim -8}\kappa\sin\delta
{T_{\rm RH}^{\tilde\phi}\over M_{\rm susy}}.
\label{eqz}
\end{equation}
It should be noted that this result is independent of the value of
$\lambda_{\rm eff}$.
The period of $\tilde\phi$ decay depends on the power $r$ in $W_2$.
In the case of $r=5$ 
where $\phi QQQQ\bar U/\tilde M_{\rm pl}^3$ is, for example,
contained in $W_2$,\footnote{We can easily introduce a discrete
symmetry to realize this by slightly modifying eq.~(\ref{eqw}).}
the decay of $\tilde\phi$ occurs sufficiently after the 
one of $\tilde\Phi_0$ and eq.~(\ref{eqx}) shows that the reheating
temperature is $T^{\tilde\phi}_{\rm RH}\sim O(10)$~GeV.
In this case the required value of $\delta$ to realize the observed
$Y_B$ should be maximal and $\kappa\sim 1$.
The $r\ge 6$ case seems to be ruled out based on the baryogenesis.

\section{Summary}
We have studied a $D$-term inflation scenario in a supersymmetric model 
extended from the MSSM by both the two abelian factor groups and several 
SM singlet chiral 
superfields. One of these groups is assumed to have a Fayet-Iliopoulos 
$D$-term $\xi$. Although the inflation scenario seems to be very similar 
to the ordinary $D$-term inflation model, our model can also have a lot of
interesting features in particle physics. The condensates of the 
additionally introduced SM singlet fields to induce the inflation can be 
directly related to both the solution of the $\mu$-problem and the 
explanation of the neutrino masses. They cause small bilinear
$R$-parity violating terms which can induce the neutrino-neutralino 
mixing. The small neutrino masses can be generated through the weak 
scale seesaw mechanism based on this mixing. The decay of such 
condensates can also induce the leptogenesis through the effective coupling
which violates the lepton number spontaneously.
Although the reheating temperature due to these decays is not so high 
that the gravitino problem can be escapable, it is high enough that 
the electroweak sphaleron interaction can transfer the generated lepton
number asymmetry into the baryon number asymmetry. 
This model can reconcile or relax the discrepancy between both values of 
$\xi$ which are required by the CMB data and by the weakly-coupled 
superstring models if the SM singlet fields relevant to the inflation
have appropriate charges of the additionally introduced abelian factor groups. 
Although there remains a lot of unstudied subjects, this kind of model
seems to have interesting features in both astrophysics and particle physics.
 
\vspace*{5mm}
This work is supported in part by a Grant-in-Aid for Scientific 
Research (C) from Japan Society for Promotion of Science
(No.~11640267 and No.~14540251) and also by a Grant-in-Aid for Scientific 
Research on Priority Areas (A) from The Ministry of Education, Science,
Sports and Culture (No.~14039205).


\end{document}